\documentstyle[epsfig]{ioplppt}       

\begin{document}
\jl{8}					

\title {Hopf's last hope: 
	spatiotemporal chaos in terms of
	unstable recurrent patterns
	}[Spatiotemporal chaos in terms of recurrent patterns]

\author{F Christiansen\ftnote{1}{
	current address: Max--Planck--Institut f\"ur Physik komplexer
Systeme, Bayreuther Strasse 40,
Haus 16, D-01187 Dresden, Germany}, 
	P Cvitanovi{\'c} and V Putkaradze}

\address
{Center for Chaos and Turbulence Studies\\
Niels Bohr Institute \\
Blegdamsvej 17, DK-2100 Copenhagen \O, Denmark}

\begin{abstract}
Spatiotemporally chaotic dynamics of a
Kuramoto-Sivashinsky system is described  
by means of an infinite hierarchy of its unstable spatiotemporally
periodic solutions. An intrinsic
parametrization of the corresponding invariant set serves 
as accurate guide to the high-dimensional dynamics, and 
 the periodic orbit theory yields  several
global averages characterizing the chaotic dynamics.
\end{abstract} 

\maketitle

\newcommand{\PC}[1]{$\footnotemark\footnotetext{PC: #1}$}
\newcommand{\VP}[1]{$\footnotemark\footnotetext{VP: #1}$}
\newcommand{\FC}[1]{$\footnotemark\footnotetext{FC: #1}$}
\newcommand{\rf}     [1] {~\cite{#1}}
\newcommand{\refref} [1] {reference~\cite{#1}}
\newcommand{\refrefs}[1] {references~\cite{#1}}
\newcommand{\refeq}  [1] {(\ref{#1})}
\newcommand{\refeqs} [2]{(\ref{#1}--\ref{#2})}
\newcommand{\reffig} [1] {figure~\ref{#1}}
\newcommand{\refFig} [1] {Figure~\ref{#1}}
\newcommand{\reftab} [1] {table~\ref{#1}}
\newcommand{\refTab} [1] {Table~\ref{#1}}
\newcommand{\beq}{\begin{equation}}
\newcommand{\continue}{\nonumber \\ }
\newcommand{\nnu}{\nonumber}
\newcommand{\eeq}{\end{equation}}
\newcommand{\ee}[1] {\label{#1} \end{equation}}
\newcommand{\bea}{\begin{eqnarray}}
\newcommand{\ceq}{\nonumber \\ & & }
\newcommand{\eea}{\end{eqnarray}}
\newcommand{\barr}{\begin{array}}
\newcommand{\earr}{\end{array}}
\renewcommand{\det}{\mbox{\rm det}}
\newcommand\period[1]{{T_{#1}}}                 
\newcommand{\prpgtr}[1]{\delta\negthinspace\left( {#1} \right)}
\newcommand{\oneMinJ}[1]{\left|\det\left({\bf 1}-{\bf J}_p^{#1}\right)\right|}
\section*{Introduction}
In recent years unstable periodic orbits have
been shown to be an effective tool in the description of 
deterministic dynamical systems of low intrinsic dimension\cite{CHAOS92}, 
in diagnosing deterministic chaos in noisy biological 
systems\cite{Moss94}, and many other applications. 
The theory has been successfully applied to low dimensional ordinary 
differential equations (deterministic chaos) and linear partial differential 
equations (semiclassical quantization). It is an open question whether
 the theory has anything to say about nonlinear partial differential 
equations (hydrodynamics, field theory).  
In this paper we show that 
 the periodic orbit theory can be used to describe 
spatially extended systems by applying it to the Kuramoto-Sivashinsky
equation\cite{Kur,Siv}.

In what follows we shall refer to a periodic solution 
 as a ``cycle'', and to the closure of the union of all 
periodic solutions as the ``invariant set''.
Periodic solutions are important because they form the {skeleton}
of the invariant set\cite{cycprl,AACI}, with
cycles ordered {hierarchically}; short cycles give good
approximations to the invariant set, longer cycles refinements.
Errors due to neglecting long cycles can be bounded, and for nice hyperbolic
systems they fall off exponentially or even superexponentially 
with the cutoff cycle length\cite{Rugh92,CRR93}.
Furthermore, cycles are {structurally robust} as for smooth flows
eigenvalues of short cycles vary slowly with smooth parameter changes,
short cycles can be accurately extracted from experimental
or numerical data, and
global averages (such as 
correlation exponents, escape rates 
and other ``thermodynamic" averages) can be efficiently computed from short
cycles by means of {cycle expansions}.

While the role of periodic solutions in elucidating the asymptotics of
ordinary differential equations was already appreciated by
Poincar\'e\cite{poincare}, allegedly Hopf\cite{Hopf42}
and, more demonstrably, Spiegel and
collaborators\cite{MS66,BMS71,EAS87} have argued that the asymptotics of
partial differential equations should also
be described in terms of recurrent spatiotemporal patterns.
Pictorially, dynamics drives a given spatially extended system through a
repertoire of unstable patterns; as we watch  
a given ``turbulent'' system evolve, 
every so often we catch a glimpse of a familiar pattern. 
For any finite  spatial resolution,
the system follows approximately for a finite time 
a pattern belonging to a finite 
alphabet of admissible patterns, and the long term dynamics can be thought
of as a walk through the space of such patterns,
just as chaotic dynamics with a  low dimensional
attractor can be thought of as a succession of nearly periodic (but
unstable) motions.

\section{Kuramoto-Sivashinsky system}

We offer here a modest implementation of the above program on
a prototype spatially extended dynamical system defined by
the Kuramoto-Sivashinsky equation\cite{Kur,Siv}
\begin{equation}
u_t=(u^2)_x-u_{xx}-\nu u_{xxxx} 
\label{ks}
\end{equation}
which arises as a model amplitude equation for interfacial instabilities
in a variety of contexts - see e.g. \refref{KNS90}.
Here $t \geq 0$ is the time, $x \in [0,2\pi]$ is the space coordinate,
and $\nu$ is a fourth-order ``viscosity'' damping parameter.
The subscripts $x$ and $t$ denote the partial derivatives with respect to 
$x$ and $t$. 
We take the Kuramoto-Sivashinsky system  because it is one of the
simplest physically interesting spatially extended nonlinear systems,
but in the present context the 
interpretation of the equation, or the equation itself is not the most 
important ingredient;
the approach should be applicable to a wide class of 
spatially extended nonlinear systems. The salient feature of such 
partial differential equations is that
for any finite value of $\nu$ their asymptotics is in principle 
describable by a
{\em finite} set of ``inertial manifold''
 ordinary differential equations\cite{Foias88}.

The program of studying unstable solutions in this context
was initiated by Goren, Eckmann and Procaccia\cite{GEP}
who have used a 2-unstable modes truncation of 
the Kuramoto-Sivashinsky equation
to study the dynamics connecting coexisting unstable 
{\em temporally stationary} solutions.
We shall study here
unstable {\em  spatiotemporally  periodic} solutions of the {\em full}
Kuramoto-Sivashinsky system.
Our main result is that in the limit of weak turbulence or 
``spatiotemporal chaos'', we can determine hierarchically and exhaustively
cycles of longer and longer periods, and apply this data 
to the evaluation of global averages.  

The function $u(x,t)=u(x+2\pi,t)$ is assumed periodic on 
the $x \in [0,2\pi]$ interval. 
As $u(x,t)$ has compact support, the standard
strategy is to expand it in a discrete spatial Fourier series: 
\begin{equation}
 u(x,t)= \sum_{k=-\infty}^{+ \infty} b_k(t) \e^{\i k x}
\, . 
\label{fseries}
\end{equation}
Since $u(x,t)$ is real, $b_k=b_{-k}^*$.
Substituting (\ref{fseries}) into (\ref{ks}) yields 
the infinite ladder of evolution equations for the Fourier coefficients $b_k$: 
\begin{equation}
\dot{b}_k=(k^2-\nu k^4)b_k +\i k \sum_{m=-\infty}^{\infty}
b_m b_{k-m}
\,. 
\label{expanfull}
\end{equation}
As  $\dot{b}_0=0$, the average (the mean drift) of the solution is 
an integral of motion. In what follows we shall assume that this average is 
zero, $\int \d x \, u(x,t) =0$.    

The coefficients $b_k$ are in general complex functions of time. 
We can simplify the system (\ref{expanfull}) further by assuming 
that $b_k$ are pure imaginary, $b_k= \i a_k$, 
where $a_k$ are real. 
As we shall see below, this picks out the
subspace of odd solutions $u(x,t)=-u(-x,t)$, with 
the evolution equations 
\begin{equation}
\dot{a}_k=(k^2-\nu k^4)a_k - k \sum_{m=-\infty}^{\infty} a_m a_{k-m} 
\,.
\label{expan}
\end{equation}
We shall determine 
the periodic solutions in the space of Fourier coefficients,
and then reconstitute from them the unstable spatiotemporally 
periodic solutions of
\refeq{ks}. 

The trivial solution $u(x,t)=0$ is  a fixed point of (\ref{ks}). From 
(\ref{expan}) it follows that the $|k|<1/ \sqrt{\nu}$ 
long wavelength modes of this fixed point 
are linearly unstable, and the 
$|k|>1/ \sqrt{\nu}$ short wavelength modes are stable.  
For $\nu > 1$,  $u(x,t)=0$ is the globally attractive stable fixed point;
starting with $\nu =1$ the solutions go through a rich sequence of
bifurcations, studied e.g. in \refref{KNS90}. 
Detailed knowledge of the parameter dependence of bifurcations sequences
is not needed for our
purposes; we shall take $\sqrt{\nu}$ sufficiently small so that the
dynamics can be spatiotemporally chaotic, but not so small that we would be
overwhelmed by too many short wavelength modes needed in order to accurately
represent the dynamics.

The growth of the unstable long wavelengths (low $|k|$) excites
the short wavelengths 
through the nonlinear term  in (\ref{expan}). 
The excitations thus transferred are dissipated by the strongly damped 
short wavelengths, and  a sort of ``chaotic equilibrium'' 
can emerge. The very short wavelengths $|k| \gg  1 / \sqrt{\nu}$ 
will remain small 
for all times, but the intermediate wavelengths of order
$|k| \sim 1 / \sqrt{\nu}$ 
will play an important role in maintaining the dynamical equilibrium. 
As the damping parameter decreases, the solutions increasingly take on 
 Burgers type shock front
character which is poorly represented by the Fourier basis, and many
higher harmonics need to be kept\cite{KNS90,GEP} in truncations of
(\ref{expan}).
Hence, while one may truncate the high modes in the expansion (\ref{expan}), 
care has to be exercised to ensure that no
modes essential to the dynamics are  chopped away.

Before proceeding with the calculations, we take into account
the symmetries of the solutions and describe our criterion for reliable
truncations of the infinite ladder of 
ordinary differential equations (\ref{expan}).

\section{Symmetry decomposition}
As usual, the first step in analysis of such dynamical flows
is to restrict the dynamics to a Poincar\'e section. We
shall fix the  Poincar\'e section to be the hyperplane
$a_1=0$. We integrate (\ref{expan}) with the initial
 conditions  
$a_1=0$, and arbitrary values of the coordinates  $a_2, \ldots, a_N$, where 
$N$ is the truncation order.  When $a_1$ becomes 
$0$ the next time,  the coordinates  $a_2, \ldots, a_N$ are mapped 
into $(a_2', \ldots a_N')=P(a_2, \ldots, a_N)$, where $P$ is the  Poincar\'e 
map. $P$ defines a mapping of a $N-1$ dimensional hyperplane into itself. 
Under successive iterations of  $P$, any trajectory
approaches the attractor ${\cal A}$, which itself is an invariant 
set under $P$. 

A trajectory of 
 (\ref{expan}) can cross the plane $a_1=0$ in two possible ways: 
 either when   
$\dot{a_1}>0$ (``up'' intersection) 
or when  $\dot{a_1}<0$ (``down'' intersection),
 with the ``down'' and ``up'' crossings 
alternating. 
It then makes sense to define the  Poincar\'e map $P$ as a transition between, 
say, ``up'' and ``up'' crossing. 
With  Poincar\'e section defined as the ``up-up'' transition, 
it is natural to define a ``down-up'' transition map $\Theta$. Since 
$\Theta$ describes the transition from down to up (or up to down) state, 
the map $\Theta^2$ describes the transition  up-down-up, that is  
$\Theta^2=P$.

Consider the spatial flip and
shift symmetry operations $Ru(x)=u(-x)$, $Su(x)=u(x+\pi)$. 
 The latter symmetry reflects the invariance under
the shift $u(x,t) \rightarrow u(x+ \pi,t)$, and is a particular case of the   
translational invariance of the Kuramoto-Sivashinsky equation (\ref{ks}). 
In the Fourier modes decomposition (\ref{expan}) this 
symmetry acts as
$S: a_{2k} \rightarrow a_{2k}, a_{2k+1} \rightarrow -a_{2k+1}$.  
Relations $R^2=S^2=1$
induce decomposition of the space of solutions into 4 invariant 
subspaces\cite{KNS90};
the above restriction to $b_k= \i a_k$ amounts to specializing
to a subspace of odd solutions $u(x,t)=-u(-x,t)$.

Now, with the help of the symmetry $S$ 
the  whole attractor ${\cal A}_{tot}$ can be 
decomposed into two  pieces: ${\cal A}_{tot}={\cal A}_0 \cup S 
{\cal A}_0 $ for some set ${\cal A}_0$.
 It can happen that the set ${\cal A}_0$
 (the symmetrically decomposed attractor)  
can be decomposed  even further by the action of the map $\Theta$. In this 
case the attractor will consist of four disjoint sets: 
 ${\cal A}_{tot}={\cal A} \cup S {\cal A}  \cup \Theta {\cal A} 
  \cup \Theta S {\cal A} $. As we shall see, 
this decomposition 
is not always possible, since sometimes $ {\cal A}$ overlaps with 
$\Theta S{\cal A} $ (in this case $\Theta  {\cal A}$ will also  overlap with 
$S {\cal A} $). 
We shall carry out our calculations in the regime where 
the decomposition into four disjoint pieces 
is possible. In this case  the set $ {\cal A}$ can be taken as
the fundamental 
domain of the Poincar{\'e} map, with $S  {\cal A} $, 
$\Theta  {\cal A} $ and $\Theta S  {\cal A} $ its images under the 
$S$ and $\Theta$ mappings.

This reduction of the dynamics to the fundamental domain is particularly
useful in periodic orbit calculations, as it simplifies symbolic dynamics
and improves the convergence of cycle expansions\cite{CEsym}.

\section{Fourier modes truncations}
 
When we simulate the equation (\ref{expan}) on a computer, we  have 
to  truncate the ladder of equations to a finite length $N$, i.e., set 
 $a_k=0$ for $k>N$. 
$N$ has to be sufficiently large that no harmonics 
$a_k$ important for the dynamics  with $k>N$ 
are truncated. On the other hand,
computation time increases dramatically with the increase of $N$:
since we will be evaluating the stability matrices for the flow,
the computation time will grow at least as $N^2$.

Adding an extra dimension to a truncation of the system (\ref{expan})
introduces a small
perturbation, and this can (and often will) 
throw the system into a totally different asymptotic state. 
A  chaotic attractor for $N=15$ can become a period three 
window for $N=16$, and so on. 
If we compute, for example, the Lyapunov exponent
$\lambda(\nu,N)$ for the strange attractor of the 
system (\ref{expan}), there is no reason to 
expect $\lambda(\nu,N)$ to smoothly converge to the limit  
value $\lambda(\nu,\infty)$ as $N \rightarrow \infty$. 
The situation is different in the periodic windows, 
where the system is structurally stable, and it makes sense to compute 
 Lyapunov exponents, escape rates, etc. for the 
{\em repeller}, i.e. the closure of the set of all 
{\em unstable} periodic orbits. 
Here the power of cycle expansions comes in: 
to compute quantities on the repeller by direct averaging methods is 
generally more difficult, because the motion quickly collapses to the 
stable cycle. 

\begin{figure}
\centerline{\epsfig{file=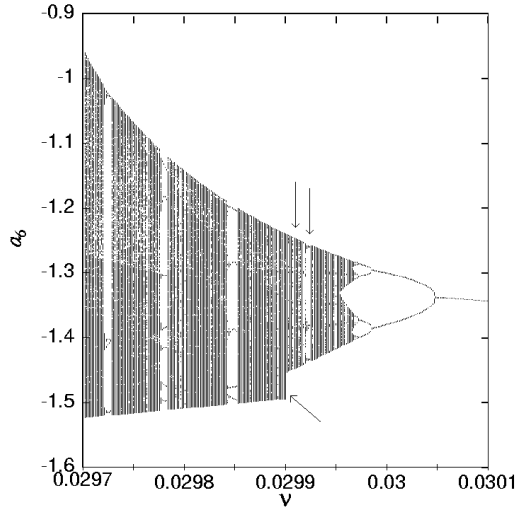,width=8cm}}
\caption[]{
Feigenbaum tree for coordinate $a_6$, $N=16$ Fourier modes 
truncation of (\ref{expan}). The two upper arrows indicate 
the values of damping parameter that we use in our 
numerical investigations; $\nu=0.029910$ 
(chaotic) and $\nu=0.029924$ (period-3 window). The lower arrow indicates
the kink where the invariant set ${\cal A}$ 
starts to overlap with $\Theta S {\cal A}$.
Truncation to $N=17$ modes yields a similar figure, with values for 
specific bifurcation points shifted by $\sim 10^{-5}$ with respect to the 
$N=16$ values. The choice of the coordinate $a_6$ is arbitrary;
projected down to any coordinate, the tree is qualitatively the same.
}
\label{feig16}
\end{figure}

We have found that the minimum value of $N$ to get any chaotic behavior at all 
was $N=9$. However, the dynamics for the $N=9$ truncated system is rather
different from the full system dynamics, and therefore we have performed our
numerical calculations for $N=15$, $N=16$ and $N=17$. 
\refFig{feig16} is a representative plot of  the Feigenbaum 
tree for the Poincar\'e map $P$. 
To obtain this figure, we took a random 
initial point, iterated it for a some  time to let it settle on the
attractor and then plotted the $a_6$ coordinate of the next 1000 intersections
with the Poincar{\'e} section.
Repeating this for different values of the damping parameter $\nu$, one can 
obtain a picture of the attractor as a function of $\nu$.
For an intermediate range of values of
$\nu$, the dynamics exhibits a rich variety of behaviours, such
as strange attractors, stable limit cycles, and so on.
The Feigenbaum trees for different values of $N$ resemble 
each other, but the precise 
values of $\nu$ corresponding the various bifurcations 
depend  on the order of truncation $N$. 

Based on  the observed numerical similarity
between the Feigenbaum trees for $N=16$ and $N=17$ (cf. \reffig{feig16}),
 we choose $N=16$ as a reasonable cutoff 
and will use only this truncation throughout the remainder of this 
paper. We will
examine  two values of the damping parameter: $\nu=0.029910$, 
for which
the system is chaotic, and $\nu=0.029924$, for which  the system has a
stable period-3 cycle.
In our numerical work we use both the pseudospectral\cite{Laurette} and the 
$4$-th order variable-step Runge-Kutta integration routines\cite{NAG};
their results are in satisfactory  agreement. As will be seen below, the
good control of symbolic dynamics guarantees that we do not miss
any short periodic orbits generated by the bifurcation sequence indicated
by the  Feigenbaum tree of \reffig{feig16}. However, even though
we are fairly sure that for this parameter value we have all
short periodic orbits, the possibility that
other sets of periodic solutions exist somewhere else in the
phase space has not been excluded.

\begin{figure}
\centerline{\epsfig{file=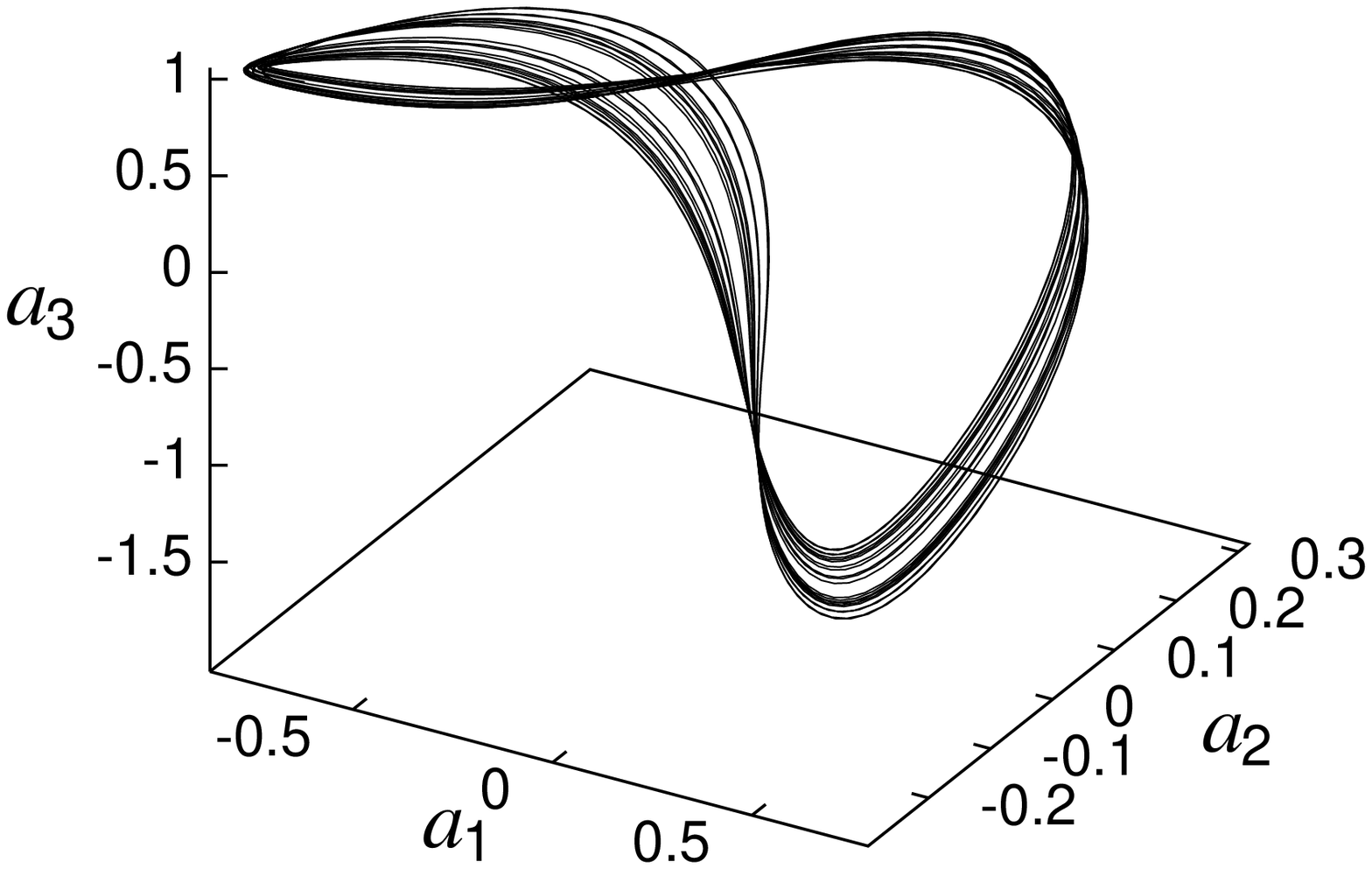,width=8cm}
	 \epsfig{file=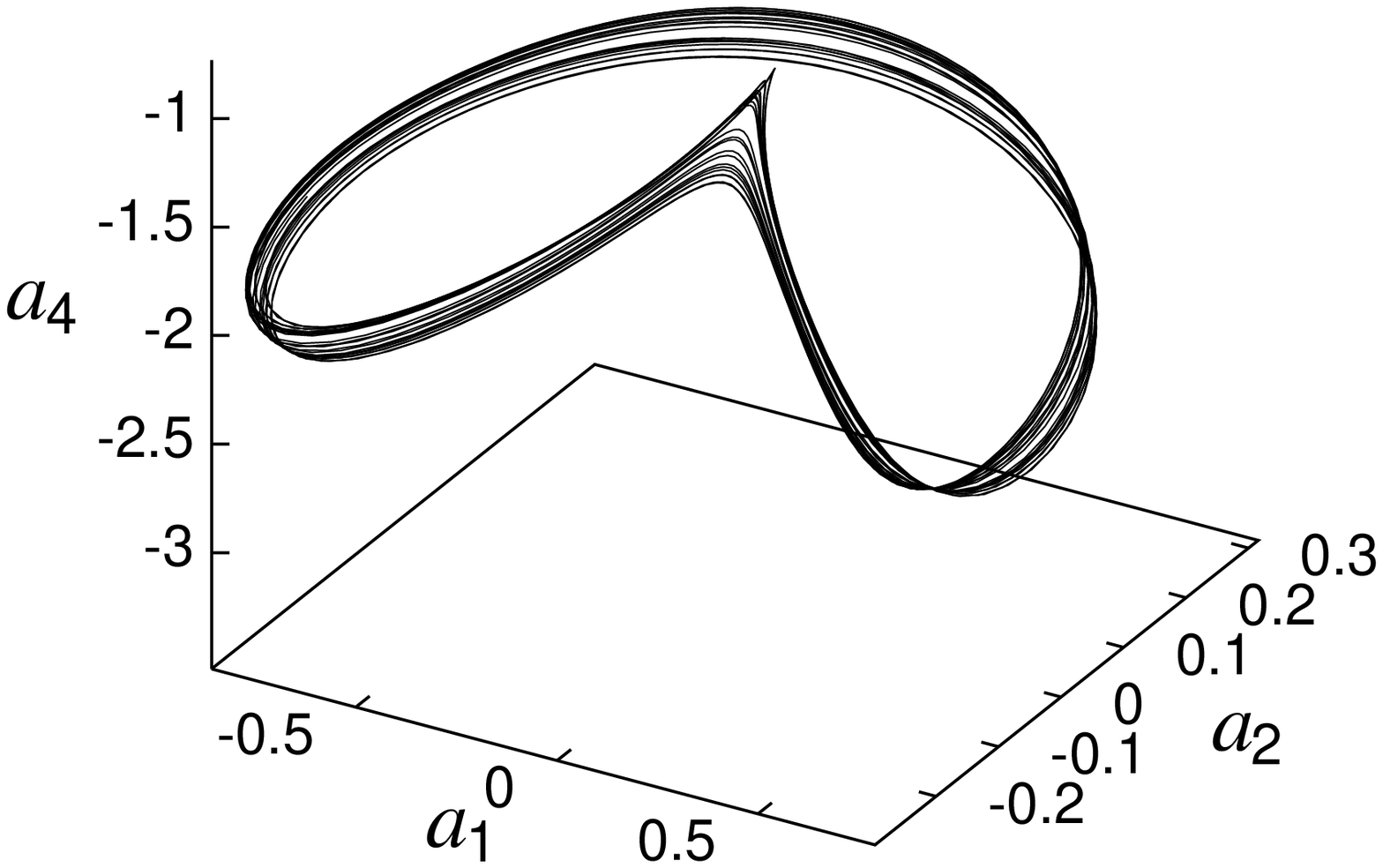,width=8cm}}
\caption[]{Projections of a typical 16-dimensional trajectory onto 
		different 3-dimensional subspaces, coordinates 
		(a) $\{a_1, a_2, a_3\}$,
		(b) $\{a_1, a_2, a_4\}$. $N=16$ Fourier modes truncation with
		    $\nu=0.029910$.}
\label{plot123}
\end{figure}

\begin{figure}
\centerline{\epsfig{file=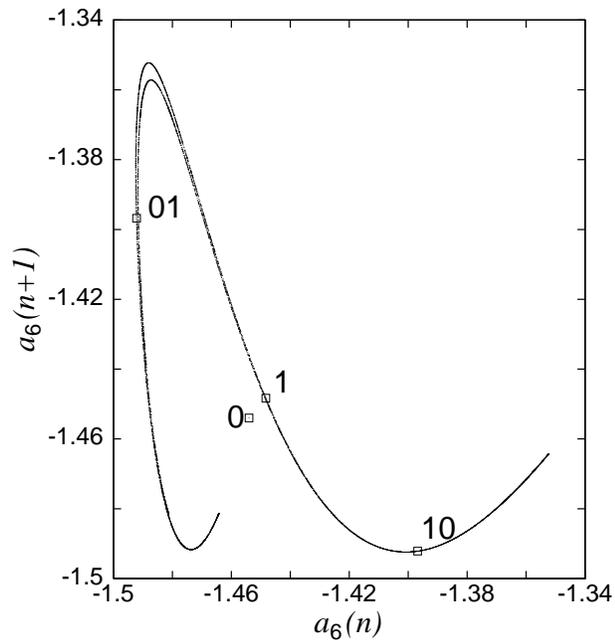,width=8cm}}
\caption[]{The attractor of the system \refeq{expan}, plotted as the $a_6$
component of the $a_1=0$  
Poincar\'e section return map, 10,000 Poincar\'e section returns of
a typical trajectory. Indicated are the periodic points $\overline{0}$, 
$\overline{1}$ and $\overline{01}$; as this is an arbitrary projection of
the invariant set, they exhibit no good spatial ordering.
$N=16$ Fourier modes truncation with $\nu=0.029910$. 
}
\label{returnmap}
\end{figure}

The problem with such high dimensional truncations of (\ref{expan})
is that the dynamics is difficult to visualize. We can 
examine its projections onto any three axes
$a_i,a_j,a_k$, as in \reffig{plot123}
or, alternatively, study 
a return map for a given  coordinate
$a_k \rightarrow a_k' = P_k(a_2, \ldots, a_{N})$
as the one  plotted in \reffig{returnmap}.
The full return map is $(N-1)$-dimensional
${\bf a} \rightarrow {\bf P}(a_2, \dots, a_{N})={\bf a}'$
and single-valued, and for the values of $\nu$ used here
the attractor is essentially 1-dimensional,
but its  projection into the $\{a_k,P_k(a_2,\ldots,a_{N})\}$ plane 
can be  multi-valued and self-intersecting.
One can  imagine a situation where no
``good'' projection is possible,
that is, any projection onto any two-dimensional
plane is a multiple-valued function.
The question is how to treat such a map?

\section{One-dimensional visualization of  the dynamics} 

We now describe an 
approach which simplifies matters a lot by reducing the 
map to an approximate one-dimensional map. The multiple-valuedness in 
\reffig{returnmap} arises from the fact that the return map 
is a 2-dimensional
projection of a convoluted 
1-dimensional curve embedded into a high-dimensional space.
We shall show that it is possible to find an {\em intrinsic} 
parametrization $s$ along the unstable manifold, 
such that the map $s \rightarrow f(s)$ induced by the
full $d$-dimensional flow is approximately {\em   $1$-dimensional}. 
Strictly speaking, the attractor on \reffig{returnmap} has a certain 
thickness transverse to it, but the contraction in the transverse
directions is so strong that the invariant set is effectively
$1$-dimensional. 

Suppose we already have determined some of the shorter cycles for our
system, i.e. 
the fixed points of the Poincar\'e map and its iterates. This is 
accomplished relatively easily 
by checking a trajectory of a random initial point for close returns
and then using these as initial guesses for a cycle search algorithm.
We now assume that the invariant set can be approximated by a curve
passing close to all periodic points,
and determine the order of periodic points along such curve.
This is done in the following way: there exists a fixed point which
is not connected to the attractor (the point 
$\overline{0}$ in \reffig{returnmap}) - we
choose this fixed point 
as the starting point and assign it number $1$. 
Point number $2$ is the periodic point in the sample
which is closest (in the full space) 
to this fixed point, and the $n$-th point is determined as the point 
which has the minimum distance from the point number $n-1$ among 
all the periodic points which have not yet been enumerated. 
Proceeding this way, we order all the periodic points that we have found
so far.

Since all periodic points belong to cycles, 
their images are known and are simply the successive periodic points
along the cycle. We use this fact to recursively construct
a $1$-dimensional mapping $s_i \rightarrow f(s_i)$. We approximate
parametrization length $s$ along the invariant set 
by computing the Euclidean inter-distances between the
successive periodic points in the full dynamical
space,
$s_1=0, s_2=\| {\bf a}_2-{\bf a}_1 \|, s_i-s_{i-1}=\| 
{\bf a}_i-{\bf a}_{i-1} \| $. 
The $i$-th cycle point $s_i$ is mapped onto its image
$s_{\sigma{i}} = f(s_i)$, 
where $\sigma i$ denotes the label of the next periodic point 
in the cycle.
We can now find longer 
periodic orbits of the 1-dimensional map $f$  by standard methods  such
as inverse iteration, and guess the location of 
the corresponding points in the full $N$-dimensional
space by interpolating between the nearest known periodic points.
These will not be exact periodic orbits of the full system,
but are  very useful as good starting guesses in a search for the exact
periodic orbits. Iteratively, more and more periodic orbits can be computed. 
While it only pays to refine the 1-dimensional parametrization 
until the density
of the periodic points become so high that the width of the attractor
becomes noticeable, the 1-dimensional map continues to provide good
initial guesses to longer periodic orbits.
More sophisticated methods are needed only if 
high accuracy around the folding region of $f(s)$ is required
in order to distinguish between long cycles.

\begin{figure}
\centerline{\epsfig{file=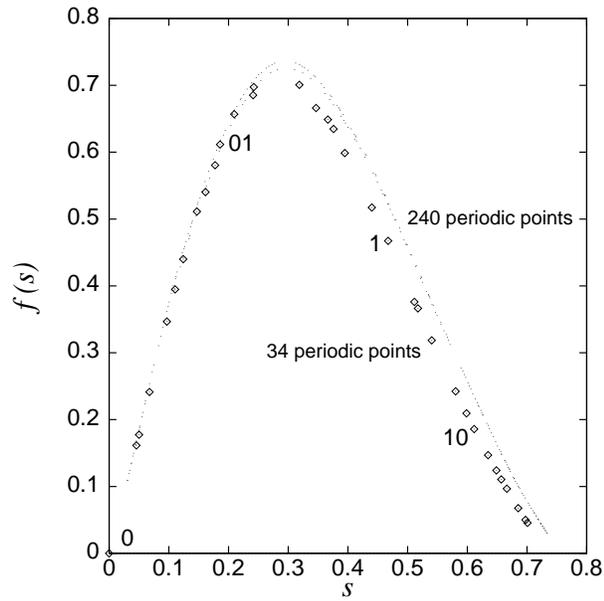,width=8cm}}
\caption[]{The return map $s_{n+1} = f(s_n)$ constructed
from the images of periodic points. The diamonds were obtained by 
using $34$ periodic points, the dots were obtained by using $240$ periodic 
points. 
We have indicated the periodic points $\overline{0}$, 
$\overline{1}$ and $\overline{01}$. 
Note that the transverse fractal structure of the map shows when 
the number of points is increased. 
$N=16$ Fourier modes truncation with $\nu=0.029910$. }
\label{unfolded}
\end{figure}
For the values of $\nu$ we are working with, the attractor consists 
of four disjoint sets, the fundamental domain  ${\cal A}$ and 
its images under the maps $S$ and $\Theta$. 
In this case the approximate return map 
$s \rightarrow f(s)$ is unimodal.
The corresponding map on the symmetric 
part of the attractor, $S \Theta {\cal A} $,  is likewise unimodal,
and turned $180$ degrees around the origin. For the values of $\nu$ we
work with the
two maps do not interact and their domains are separate. 
However, if the value of the damping parameter $\nu$ is decreased
sufficiently, the domains of the
maps join and together they form a connected invariant set
described by a bimodal map\cite{Glbtsky94}.
This joining of the fundamental domain ${\cal A}$ and 
its symmetric image $\Theta S {\cal A} $ is visible 
in \reffig{feig16} at
$\nu \simeq 0.0299$, where the range of the $a_6$ coordinate 
increases discontinuously.

\begin{figure}
\centerline{\epsfig{file=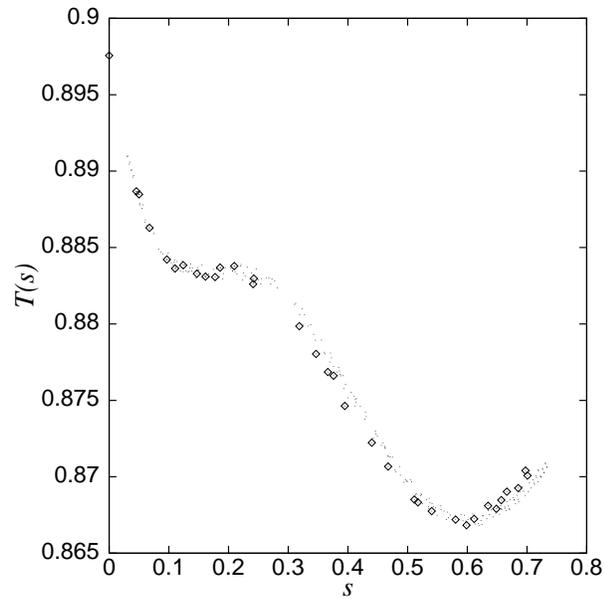,width=8cm}}
\caption[]{The return time $T(s)$ as a function of the parameter $s$,
evaluated on the periodic points, as in \reffig{unfolded}, with
the diamonds obtained by 
$34$ periodic points and the dots by $240$ periodic points. 
The fine structure is due to the fractal structure of the attractor.
}
\label{rettime}
\end{figure}

We use the unimodal map $s \rightarrow f(s)$ to construct binary symbolic 
dynamics for the system in the usual way: assign the symbol '0' to points
to the left of the maximum, '1' to the points to the right. 
In the period-3 window with the stable cycle $\overline{001}$, 
the pruning rules are very easy: except for the stable $\overline{001}$ cycle
and the $\overline{0}$ fixed point (both disjoint from the invariant set) 
two 0's in a row are forbidden. 
In this case it is convenient to redefine the alphabet by
denoting the symbol pair $01$ by $a$ and the symbol $1$ by $b$.
This renders the symbolic dynamics of the points on the repeller
complete binary: all sequences of
the letters $a$ and $b$ are admissible.

A flow in $N$ dimensions 
can be reduced to an $(N-1)$-dimensional 
return map by suspension on a Poincar\'e section
provided that the Poincar\'e return map is 
supplemented by a ``time ceiling function''\cite{bowen} which accounts
for a variation in the section return times. 
Hence we also determine the return time $T(s_i)$ 
for each periodic point $i$, and use those to construct recursively
the periodic orbit approximations to the time ceiling function,
\reffig{rettime}.
The mean Poincar\'e section return time 
is of order $\overline{T} \approx .88$.

\subsection{Numerical results}

We have found all cycles 
up to topological length 10 (the redefined topological length in the case of
the period-3 window), 92 cycles in the chaotic regime and 228 
in the  period-3 window,
by using the 1-dimensional parametrization $f(s)$ to find initial guesses for
periodic points of the full $N=16$ Fourier modes truncation 
and then determining the cycles by
a multi-shooting Newton routine. It is worth
noting that the effectiveness of using the 
$1$-dimensional $f(s)$ approximation to the dynamics
to determine initial guesses is
such that for a typical cycle it takes only 2-3 Newton iterations to find
the cycle with an accuracy of $10^{-10}$.

\begin{table}
\caption[]{All cycles up to topological length
5 for the $N=16$ Fourier modes truncation of the Kuramoto-Sivashinsky equation 
(\ref{expan}),
damping parameter 
$ \nu =0.029910$ (chaotic attractor) and $\nu=0.029924$ (period-3
window), their itineraries, periods, 
the first four stability eigenvalues. For the chaotic attractor
pruning shows up at the 
topological length $4$; $\overline{0001}$ 
and $\overline{0011}$ cycles are pruned. 
The deviation from unity of $\Lambda_2$, the eigenvalue along the flow, 
is an indication of the accuracy of the numerical integration.  
For the period-3 window we also give the itineraries in the redefined alphabet
where $a=1$ and $b=10$.}
\begin{indented}
{\small 
\lineup
\item[]\begin{tabular}{@{}lllllll}
\br
$p$ & & $T_p$ & $\Lambda_1$ & $\Lambda_2-1$ & $\Lambda_3$ & $\Lambda_4$ \\ \mr
\multicolumn{7}{l}{Chaotic, $\nu=0.029910$} \\ \mr
0 & & 0.897653 & \03.298183 & 5$\cdot10^{-12}$ 
& \-2.793085$\cdot 10^{-3}$ & \-2.793085$\cdot 10^{-3}$ \\
1 & & 0.870729 & \0\-2.014326 & \-5$\cdot10^{-12}$ 
 & 6.579608$\cdot 10^{-3}$ & \-3.653655$\cdot 10^{-4}$ \\
10 & & 1.751810 & \0\-3.801854 & 8$\cdot10^{-12}$ &
 \-3.892045$\cdot 10^{-5}$ & 2.576621$\cdot 10^{-7}$ \\
100 & & 2.639954 &\0\-4.852486 & 1$\cdot10^{-11}$
 & 3.044730$\cdot 10^{-7}$ &\-3.297996$\cdot 10^{-10}$ \\
110 & & 2.632544 & \06.062332 & 2$\cdot10^{-11}$ &
 \-2.721273$\cdot 10^{-7}$ & \-1.961928$\cdot 10^{-10}$ \\ 
1000 & & - & - & - & - & - \\
1100 & & - & - & - & - & - \\
1110 & &  3.497622 & \-14.76756 & 2$\cdot10^{-11}$ &
 \-1.629532$\cdot 10^{-9}$ & 6.041192$\cdot 10^{-14}$ \\
10100 & & 4.393973 & 19.64397 & 2$\cdot10^{-11}$ & 
\-1.083266$\cdot 10^{-11}$ & 3.796396$\cdot 10^{-15}$ \\
11100 & & 4.391976 & \-18.93979 & 2$\cdot10^{-11}$ & 
 1.162713$\cdot 10^{-11}$ &\-1.247149$\cdot 10^{-14}$ \\
11010 & & 4.380100 & \-26.11626 & 2$\cdot10^{-11}$ & 
 1.005397$\cdot 10^{-11}$ & 8.161650$\cdot 10^{-15}$ \\
11110 & & 4.370895 & 28.53133 & 2$\cdot10^{-11}$ & 
1.706568$\cdot 10^{-11}$ & 1.706568$\cdot 10^{-14}$ \\ \mr
\multicolumn{7}{l}{Period-3 window, $\nu=0.029924$} \\ \mr
     0 & & 0.897809 &  \03.185997 & 7$\cdot10^{-13}$ & 
\-2.772435$\cdot10^{-3}$ & \-2.772435$\cdot10^{-3}$ \\ 
     1 & $a$ & 0.871737 & \0\-1.914257 & 5$\cdot10^{-13}$ &
 6.913449$\cdot10^{-3}$ & \-3.676167$\cdot10^{-4}$ \\ 
    10 & $b$ & 1.752821 & \0\-3.250080 & 1$\cdot10^{-12}$ &
\-4.563478$\cdot10^{-5}$ &  2.468647$\cdot10^{-7}$ \\ 
   100 & & 2.638794 & \0\-0.315134 & \-4$\cdot10^{-13}$ &
 4.821809$\cdot10^{-6}$ & \-2.576341$\cdot10^{-10}$ \\ 
   110 & $ab$ & 2.636903 &  \02.263744 & 3$\cdot10^{-12}$ &
\-6.923648$\cdot10^{-7}$ & \-2.251226$\cdot10^{-10}$ \\ 
  1110 & $aab$ & 3.500743 & \-10.87103 & 2$\cdot10^{-12}$ &
\-2.198314$\cdot10^{-9}$ &  3.302367$\cdot10^{-14}$ \\ 
 11010 & $abb$ & 4.382927 & \-15.84102 & 2$\cdot10^{-12}$ &
 1.656690$\cdot10^{-11}$ &  1.388232$\cdot10^{-14}$ \\ 
 11110 & $aaab$ & 4.375712 &  18.52766 & 3$\cdot10^{-12}$ &
\-1.604898$\cdot10^{-11}$ &  2.831886$\cdot10^{-14}$ \\ \br
\end{tabular} 
}
\end{indented}
\label{t_orbits}
\end{table}

\begin{figure}
\centerline{\epsfig{file=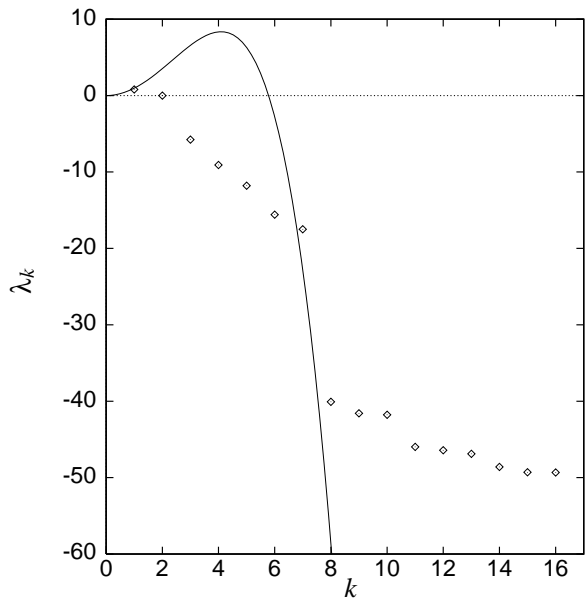,width=8cm}}
\caption[]{Lyapunov exponents $\lambda_k$ versus $k$ for the 
periodic orbit $\overline{1}$ compared with  the stability eigenvalues 
of the $u(x,t)=0$ stationary solution $k^2- \nu k^4$.  
$\lambda_k$ for $k \geq 8$ fall below the numerical accuracy of integration 
and are not meaningful. 
$N=16$ Fourier modes, $\nu=0.029924$, chaotic regime.
}
\label{eigenvalues}
\end{figure}

In \reftab{t_orbits} we list the periodic orbits 
to topological length 5 found by our method.
The value of $\Lambda_2$ serves as an indication of the 
accuracy of our numerics, as $\Lambda_2$ corresponds to the marginal
eigenvalue along the periodic orbit, strictly equal to $1$.
All cycles seem to have 
real eigenvalues (to within the numerical accuracy) except for the 
$\overline{0}$-cycle 
which has a pair of complex eigenvalues, $\Lambda_3$ and $\Lambda_4$.
We therefore do not list the corresponding imaginary parts of the eigenvalues. 
To illustrate the rapid contraction in the nonleading eigendirections
we plot all the eigenvalues of the 
$\overline{1}$-cycle in \reffig{eigenvalues}. 
As the length of the orbit increases, the magnitude of
contracting eigenvalues falls quickly 
bellow the attainable numerical numerical 
accuracy $\approx 10^{-16}$ and our numerical results for 
$\Lambda_k$ are not meaningful for $ k \geq 8$.

\begin{figure}
\centerline{\epsfig{file=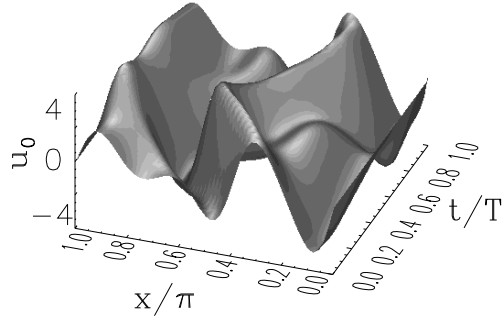,width=8cm}}
\caption[]{Spatiotemporally periodic solution $u_0(x,t)$.
We have divided $x$ by $\pi$ and plotted only the  $x>0$ part, since we work in
the subspace of the odd solutions,  $u(x,t)=-u(-x,t)$. 
$N=16$ Fourier modes truncation with $\nu=0.029910$.
	  }
\label{orbit0fig}
\end{figure}

\begin{figure}
\centerline{\epsfig{file=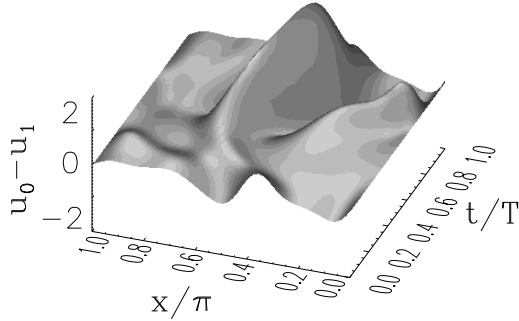,width=8cm}}
\caption[]{The difference between the two shortest period
	   spatiotemporally periodic solutions 
	   $u_0(x,t'T_0)$ and $u_1(x,t''T_1)$.
	  }
\label{diff1fig}
\end{figure}

\begin{figure}
\centerline{\epsfig{file=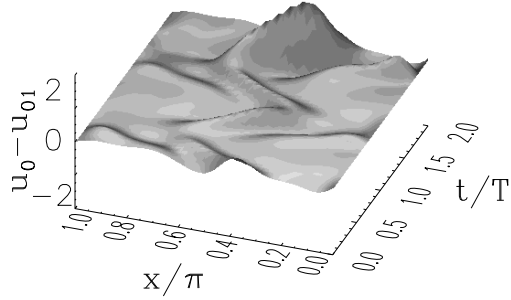,width=8cm}}
\caption[]{The difference between solution  $u_0(x,t'T_0)$ repeated
	   twice and the $n_p=2$ period spatiotemporally periodic solution
           $u_{01}(x,t''T_{01}/2)$.
          }
\label{diff01fig}
\end{figure}

Having determined the periodic solutions $p$ in the Fourier modes space, 
we now go back to the configuration space and plot the corresponding
spatiotemporally periodic solutions $u_p(x,t)$: they are the
repertoire of the recurrent spatiotemporal patterns that Hopf wanted to
see in turbulent dynamics.
Different spatiotemporally periodic solutions are qualitatively
very much alike but still different, as a closer inspection reveals. 
In \reffig{orbit0fig} we plot 
$u_0(x,t)$ corresponding to the Fourier space $\overline{0}$-cycle. 
Other solutions, plotted in the configuration space, exhibit the same
overall gross structure. For this reason we find it more informative
to plot the difference $u_0(x,t'T_0)-u_p(x,t''T_p/n_p)$
rather than $u_p(x,t)$ itself. 
Here $p$  labels a given prime (non-repeating) cycle, 
$n_p$ is the topological cycle length, $T_p$ its period,
and the time is rescaled to make this difference periodic in 
time: $t'=t /T_0$ and $t''=n_p t/T_p$, so that $t''$ ranges from $0$ to $n_p$. 
$u_0(x,t'T_0)-u_1(x,t''T_1)$ is given in \reffig{diff1fig}, and
$u_0(x,t'T_0)-u_{01}(x,t''T_{01}/2)$ in \reffig{diff01fig}.

\section{Global averaging: periodic orbits in action} 

The above investigation of the Kuramoto-Sivashinsky
system demonstrates that it is possible to
construct recursively and exhaustively
a hierarchy of spatiotemporally periodic unstable solutions
of a spatially extended nonlinear system.

Now we turn to the central issue of this paper; qualitatively, these
solutions are indeed an implementation of Hopf's program, but how
is this information to be used quantitatively? This is precisely
what the periodic orbit theory is about;
it offers machinery that puts together
the topological and the quantitative information about individual
solutions, such as their periods and stabilities, into predictions 
about measurable global averages, such as the Lyapunov exponents,
correlation functions, and so on. The proper tool for
computing such global characterizations of the dynamics are the trace
and determinant formulas of the periodic orbit theory.

We shall briefly summarize the aspects of the
periodic orbit theory relevant to the present application;
for a complete exposition of the theory 
the reader is referred to \refref{cycl_book}. The key idea is to
replace a time average $\Phi^t (x)/t$
of an ``observable"  $\phi$ measured along
a dynamical trajectory $x(t) = f^t(x)$
\[
\Phi^t (x) = \int_0^t \d\tau \, \phi(x(\tau))
\]
by the spatial average $\left<\e^{\beta \Phi^t}\right>$
of the quantity $\e^{\beta \Phi^t (x)}$. Here $\beta$ is
a dummy variable, used to recover the desired expectation value
$\left<\phi\right> = \lim_{t\to\infty} \left<\Phi^t/t \right>$ 
by taking $\frac{\d}{\d\beta}$ derivatives
of $\left<\e^{\beta \Phi^t}\right>$ and then setting $\beta=0$.
For large $t$ the average $\left<\e^{\beta \Phi^t}\right>$ behaves as
the trace 
\beq
\tr {\cal L}^{t}
     =   \sum_{p} \period{p} \sum_{r=1}^\infty
                { \e^{r \beta \Phi_p} \over \oneMinJ{r} }
                \prpgtr{t-r \period{p}}
\ee{tr-L1}
of the evolution operator
\[
{\cal L}^t (x,y) = \delta(y-f^{t}(x))\e^{\beta \Phi^t(x)} .
\]
and is dominated by its largest eigenvalue $\e^{ts(\beta)}$.

The trace formula \refeq{tr-L1} has an intuitive geometrical interpretation.
The sums in \refeq{tr-L1}
are over prime periodic orbits $p$ 
and their repeats $r$, $T_p$ are their periods,
and ${\bf J_p}$ are their stability matrices. 
Prime cycles partition the dynamical space into closed tubes of
length $\period{p}$ and thickness
$\left|\det({\bf 1}-{\bf J}_p)\right|^{-1}$, 
and the trace picks up a periodic orbit contribution only when the
time $t$ equals a prime period or its repeat, hence the time
delta function $\prpgtr{t- r\period{p}}$. Finally,
$\e^{\beta \Phi_p }$
is the mean value of
$ \e^{\beta \Phi^t(x)}$
evaluated on this part of dynamical space, so the trace formula is 
the average of $\left< \e^{\beta \Phi^t}\right>$ expressed as a partition
of the space of solutions into a repertoire of spatiotemporally
periodic solutions, each weighted by its stability, i.e. likelihood of its
occurrence in a long time evolution of the system.

In applications of the periodic orbit theory 
the related Fredholm determinant
\begin{equation}
F(\beta,s)=
\exp \left ( 
{ \displaystyle - \sum_p \sum_{r=1}^{\infty} z ^{n_p r} 
\frac{\displaystyle \e^{r( \beta \Phi_p -s T_p) } }
{ \displaystyle 
r \left | \det \left ( {\bf 1}- {\bf J}_p^r \right )  \right | } } \right )
\label{fredholm} 
\end{equation}
has better convergence as a function of
the maximal cycle length truncation, so that is the function whose
leading zero $F(\beta,s(\beta))=0$ we determine here in order
to evaluate the leading eigenvalue $s(\beta)$. 

The dummy variable $z$ in \refeq{fredholm} keeps track of 
the topological lengths $n_p$ (number of the Poincar\'e section crossings),
and is used to expand $F$ as a series in
$z$. If we know all cycles up to topological length $l$ we truncate $F$
to $l$-th order polynomial:
\begin{equation}
F(\beta,s)=1-\sum_{1}^{l} c_k z^k + (\mbox{remainder})
\label{Fred_exp}
\end{equation}
and set $z=1$. The general theory\cite{Ruelle,Rugh92,CRR93} then 
guarantees that
for a hyperbolic dynamical system the coefficients $c_k$ fall
off in magnitude exponentially or faster with increasing $k$.
We now calculate the leading eigenvalue $s(\beta)$ 
by determining the smallest zero of $F(\beta,s)$,
and check the convergence of this estimate by studying it as a 
function of the maximal cycle length truncation $l$. 
If the flow conserves all trajectories, the leading eigenvalue
must satisfy $s(0)=0$; if the invariant set is repelling, the
leading eigenvalue yields $\gamma= -s(0)$, the escape rate from the repeller.
Once the leading eigenvalue is determined we can
calculate the desired average $\left<\phi \right>$ using formula\cite{AACI}:
\begin{equation}
\left<\phi \right> =
\left. -\frac{\partial s}{\partial \beta}\right|_{\beta=0} =
\left. -\frac{\partial F}{\partial \beta} 
  \left/
 \frac{ \partial F}{\partial s  }
  \right.  \right|_{\beta=0 \atop s=s(0)}.
\label{cyc_aver}
\end{equation}
For example, if we take as our ``observable'' $\log |\Lambda_{1}^t(x)|$,
the largest eigenvalue of the linearized stability of the flow,
$\Phi_p$ will be
$\log |\Lambda_{1,p}|$ where $\Lambda_{1,p}$ is the largest eigenvalue
of stability matrix of the cycle $p$, and the above 
formula yields the Lyapunov exponent $\left<\lambda\right>$.
 
Both the numerator and the denominator in \refeq{cyc_aver} have a cycle
expansion analogous to \refeq{Fred_exp} (cf. \refref{cycl_book}), 
and the same periodic orbit data suffices for their evaluation.

Conceptually the most
important lesson of the periodic orbit theory 
is that the spatiotemporally periodic
solutions are {\em not} to be thought of as eigenmodes to be used as a linear
basis for expressing solutions of the equations of motion - as the equations
are nonlinear, the periodic solutions are in no sense additive.
Nevertheless, the trace formulas and determinants of the periodic
orbit theory give a precise prescription for how 
to systematically explore the repertoire of admissible spatiotemporal patterns,
and how to put them together in order to predict measurable observables. 

\subsection{Numerical results}

One of the objectives of a theory of turbulence is
to predict measurable global averages over turbulent flows, such as
velocity-velocity correlations and transport coefficients. While in
principle the periodic orbit averaging formulas should be applicable
to such averages, with the present parameter values
we are far from any strongly turbulent regime, and here we shall
restrict ourselves to the simplest tests of chaotic dynamics: we shall
test the theory by evaluating Lyapunov exponents
and escape rates.

\begin{figure}
\centerline{\epsfig{file=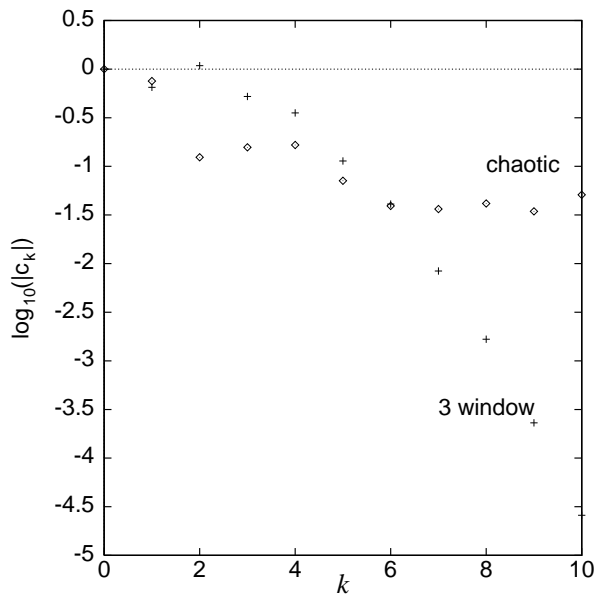,width=8cm}}
\caption[]{$\log_{10}$ of the coefficients  $|c_k|$ in the cycle expansion 
\refeq{Fred_exp} of $F(0,0)$ versus $k$ 
for the period-$3$ window case (crosses) and the chaotic case
(diamonds). $N=16$ Fourier modes truncation. }
\label{coeffval}
\end{figure}

We compute the periodic orbits, escape rates and  Lyapunov exponents 
both for the period-$3$ window and a chaotic regime.
In the case of period-$3$ window the complete symbolics dynamics
and grammar rules
are known and  good convergence of cycle expansions is expected 
both for the escape rate from the repeller
and the Lyapunov exponent. Parenthetically, the stable
period-3 orbit is separated from the rest of the invariant set
by its immediate basin of attraction window, and its eigenvalues 
bear no immediate relation to the escape rate and the Lyapunov
exponent of the repelling set.

In the case of a generic ``strange attractor'',
the convergence is not expected to be nearly as good, 
since in this case there exist no finite description of the symbolic dynamics. 
For closed systems (no escape) $\gamma=0$ and $F(0,0)=0$. 
The discrepancy of the value $F(0,0)$ from $0$ for a closed system 
allows us to estimate the accuracy of 
finite cycle length approximations to the 
Fredholm determinant. 

The analytic properties of the Fredholm determinant are illustrated by 
the decay rate of the coefficients $c_k$ as a function of $k$ in the expansion 
\refeq{Fred_exp}.
If the complete symbolic dynamics is known and the system is hyperbolic, the 
decay of $c_k$ should be superexponential\cite{Rugh92}. 
This is illustrated in \reffig{coeffval}, where
we plot the coefficients $c_k$ for the $16$-dimensional 
system for the chaotic case and for the period-$3$ window. 
We can clearly see the 
superexponential decay for the period-$3$ 
window case and at best exponential decay 
for the chaotic case. 

Our results are presented in \reftab{t_16_chaotic}. One 
observes that when the symbolic dynamics is known 
(period-$3$ window), the convergence is much better than
in the generic case, in accordance with the periodic orbit theory
expectations. 

\begin{table}
\caption[]{
The escape rate $\gamma$ and the leading Lyapunov exponent
as a function of the cycle expansion truncation $n_{max}$ 
for the $N=16$ Fourier modes truncation, chaotic regime 
$(\nu=0.029910$) and period-3 window $(\nu=0.029924)$. In the 
period-3 window the Fredholm determinant starts converging only
for $n_{max}>4$; for $n_{max}=4$ it has no real zero at all.
A numerical simulation 
estimate for the Lyapunov exponent in the chaotic regime is given in the 
last line; for this parameter value the escape rate, $\gamma$,
should strictly equal zero.}
\begin{indented}
{\small 
\lineup
\item[]\begin{tabular}{lllll} \br
 & \multicolumn{2}{l}{chaotic} & 
    \multicolumn{2}{l}{period-3 window} \\ \mr
$n_{max}$ & $\gamma$ & $\lambda_1$ & $\gamma$ & $\lambda_1$ \\ \mr
 1 & & & 0.428143 & 0.703010 \\ \hline
 2 & 0.441948 & 0.981267 &  \-0.187882 & 0.430485 \\ \hline
 3 & 0.080117 & 0.765050 &  \-0.049325 & 0.469350 \\ \hline
 4 & 0.148583 & 0.703072 & & \\ \hline
 5 & 0.068513 & 0.727498 &  1.072468 & 0.585506 \\ \hline
 6 & 0.027724 & 0.699907 &  0.078008 & 0.547005 \\ \hline
 7 & 0.035137 & 0.693852 &  0.088132 & 0.598977 \\ \hline
 8 & 0.007104 & 0.675529 &  0.090425 & 0.631551 \\ \hline
 9 & 0.021066 & 0.673144 &  0.090101 & 0.618160 \\ \hline
10 & 0.007367 & 0.646233 &  0.090065 & 0.621271 \\ \hline
numer. &  & 0.629 & & \\  \br
\end{tabular}
}
\end{indented}
\label{table16_chaotic}
\label{t_16_chaotic}
\end{table}

\section{Summary}

Hopf's proposal for a theory of turbulence was, as we understand it, to think 
of turbulence as a sequence of near recurrences of a repertoire of unstable 
spatiotemporal patterns. Hopf's 
proposal is in its spirit very different from most ideas that animate
current turbulence research. 
It is distinct from
the Landau quasiperiodic picture of turbulence as a sum of 
infinite number of incommensurate frequencies, with dynamics taking place on a 
large-dimensional torus.
It is not the
Kolmogorov's 1941 homogeneous turbulence with no 
coherent structures fixing the length scale, here all the action is 
in specific coherent structures.
It is emphatically {\em not} universal; spatiotemporally periodic solutions 
are specific to the particular set of equations and boundary conditions.
And it is {\em not} probabilistic; everything is fixed by the deterministic
dynamics with no probabilistic assumptions on the velocity distributions 
or external stochastic forcing. 

Our investigation of the Kuramoto-Sivashinsky system is a
step in the direction of implementing Hopf's program. 
We have constructed 
a complete and exhaustive hierarchy of spatiotemporally periodic solutions 
of spatially extended nonlinear system and applied the periodic orbit theory 
to evaluation of global averages for such system. Conceptually the most 
important lesson of this theory is that 
the unstable spatiotemporally periodic
solutions serve to explore systematically the
repertoire of admissible spatiotemporal patterns, with the trace
and determinant formulas and their cycle expansions being the proper tools for
extraction of quantitative predictions from the periodic orbits data.

We have applied the
theory to a low dimensional attractor, not larger than the Lorenz's original 
strange attractor\cite{Lorenz}. As our aim was to solve the given equations 
accurately, we were forced to work with a high dimensional
Fourier modes truncations, and we succeeded in determining the
periodic orbits for flows of much higher dimension than in
previous applications of the periodic orbit theory. As something new, we 
have developed an intrinsic parametrization of the invariant set that
provided the key to finding the periodic orbits. 

In practice, the method 
of averaging by means of periodic orbits produced best results when 
the complete symbolic dynamics was known. 
For generic parameter values we cannot claim that the periodic
orbit approach is
computationally superior to a direct numerical simulation. 
A program to find periodic orbits up to length 10 for one value of 
the damping parameter $\nu$ requires a day of CPU on a fast workstation, 
much longer than the time used in the direct numerical
simulations. 

The parameter $\nu$ values that we work with correspond to
the weakest nontrivial ``turbulence'', and it is an open 
question to what extent the approach remains implementable as the system goes 
more turbulent. Our hope is that the unstable structures  captured so far 
can be adiabatically tracked to the ``intermediate turbulence'' regime, 
and still remain sufficiently representative of the space of admissible 
patterns to allow meaningful estimates of global averages.  
As long as no effective coordinatization of the ``inertial
manifold'' exists and we rely on the spatial Fourier decomposition, the
present approach is 
bound to fail in the ``strong turbulence'' $\nu \rightarrow 0$ limit,
where the dominant structures are Burgers-type shocks 
and truncations of the spatial Fourier modes 
expansions are increasingly uncontrollable. 

\ack
We are grateful to L. Tuckerman for patient instruction,
E.A. Spiegel,
G. Goren, 
R. Zeitek,
and 
I. Procaccia
for inspiring conversations, P. Dahlqvist for a critical
reading of an early version of the paper, and E. Bogomolny for the
catchy but all too ephemeral title for the paper.

\end{document}